# Study of the MgB$_2$ grain size role in *ex-situ* multifilamentary wires with thin filaments

A. Malagoli[1], V. Braccini, C. Bernini, G. Romano, M. Vignolo, C. Ferdeghini and M. Putti

[1] CNR-INFM LAMIA, C.so Perrone 24, I-16152 Genova, Italy

E-mail: andrea.malagoli@lamia.infm.it

**Abstract** The MgB$_2$ superconductor has already demonstrated its applicative potential, in particular for DC applications such as MRI magnets, thanks to the low costs of the raw materials and to its simple production process. However further efforts have still to be made in order to broaden its employment also towards the AC applications such as SFCL, motors, transformers. The main issues are related to the reduction of the AC losses. Some of these can be faced by obtaining multifilamentary conductors with a large number of very fine filaments and, in this context, the powders granulometry can play a crucial role.

We have prepared MgB$_2$ starting powders with different granulometries and by the *ex-situ* P.I.T method we have realized multifilamentary wires with a number of filaments up to 361 and an average size of each filament lowered down to 30 μm. In particular we have studied the relationship between grain and filament size in terms of transport properties and show that the optimization of this ratio is possible in order to obtain suitable conductors for AC industrial applications.

## 1. Introduction

Nowadays the MgB$_2$-based conductors fabricated by the Powder-In-Tube (PIT) technique can be considered not only as promising superconductors but also as an actual industrial product. In parallel to a broad research addressed to improve the critical current density ($J_C$) behaviour, a number of groups have focused their efforts on developing multifilamentary conductors – being more interesting than a monocore from an applicative point of view – following the *ex-situ* as well as the *in-situ* route [1]. On the one hand the research has been focused on the pinning issue and several doping procedures have been developed in order to improve the behaviour in magnetic field, most of them regarding the *in-situ* [2-5] even though works using the *ex-situ* do not lack [6]. On the other hand, the clear applicative prospect has prompted the development of techniques to produce multifilamentary wires [7-10] or cables [11,12]. After few years since the discovery of superconductivity in MgB$_2$, long multifilamentary strands - of the order of km - have been fabricated by Hyper Tech Research using the *in-situ* [13] and Columbus Superconductors using the *ex-situ* [9] way. The stabilized multifilamentary tape produced by Columbus Superconductors has even been successfully employed to realize a low field magnet [14] operating at 20 K for a new superconducting M.R.I. machine that is at present marketed.

However in order to make MgB$_2$ useful not only for DC but also for AC applications further conductor development and optimization are still needed, in particular to reduce the AC losses caused by magnetic hysteresis in the MgB$_2$ core, filaments coupling and eddy currents flowing through the metallic matrix. In this context the research works should be focused on multifilamentary strands with a large number of very fine filaments, twisted filaments and non-magnetic and high resistivity sheath.

In particular the issue regarding the achievement of a large number of very fine filaments (10-30 μm) seems not to have a simple solution. The *in-situ* way has been already employed to realize multifilamentary wires with a number of filaments up to 61 [15] and a filament size down to 40 μm [16,7]. However, a $J_C$ degradation has been observed with increasing the number of filaments and in particular the granulometry of the starting precursors seems to be the bottleneck in achieving homogeneous fine filaments.

Even though the *ex-situ* conductors did not reach in field performances as good as the *in-situ* after doping, they show several industrial advantages such as lesser porosity, [1] the possibility of being used to realize React & Wind coils, [17] homogeneity over long lengths, better mechanical properties, $MgB_2$ phase and granulometry control [9].

In previous works we showed the positive effect of a finer $MgB_2$ granulometry on the superconducting properties of *ex-situ* monocore tapes [18] and we investigated the feasibility of *ex-situ* wires with very thin filaments down to 30 μm [10]. In this paper, by exploiting the possibility of modifying and controlling the grain size of the starting $MgB_2$ powders, we study the relationship between the starting granulometry and the final single filament size in terms of transport properties and show that it is possible to optimize $MgB_2$ conductors for AC applications.

## 2. Experimental details

$MgB_2$ multifilamentary wires were realized by the *ex-situ* PIT method [9]. $MgB_2$ powders were prepared from commercial amorphous B (95-97% purity) and Mg (99% purity): the powders were mixed and underwent a heat treatment at 910°C in Ar. As described in our previous work [18], the granulometry of the so obtained $MgB_2$ powders was analyzed by a Scanning Electron Microscope (SEM) and an average grain size of about 1.5 μm was estimated (see figures 1a-c). A part of these powders was ball milled and $MgB_2$ with an estimated average grain size of 450 nm was obtained (see figures 1b-c). Both not-milled (NM) and milled (M) powders were used to fill two nickel tubes with an outer diameter of 21.34 mm and a thickness of 2.11 mm. After drawing to the proper size, such monofilaments were then stacked into 19- and 91- subelement arrays inside an external monel tube with an outer diameter of 21.34 mm and a thickness of 2.11 mm. After drawing and groove rolling, a 19-filaments and a 91-filaments square wire (1.7 x 1.7 mm$^2$) for each powder (NM and M) were obtained. Moreover, 19 pieces of the monofilaments were inserted into a further nickel tube and, after drawing, each 19-filaments wire was cut in further 19 elements and restacked into a monel tube and, after the cold deformation described above, two 361-filaments square wires (1.7 x 1.7 mm$^2$) were obtained. Finally by a further special deformation step the 91- and 361-filaments square wires were round shaped with a diameter of 1.7 mm.

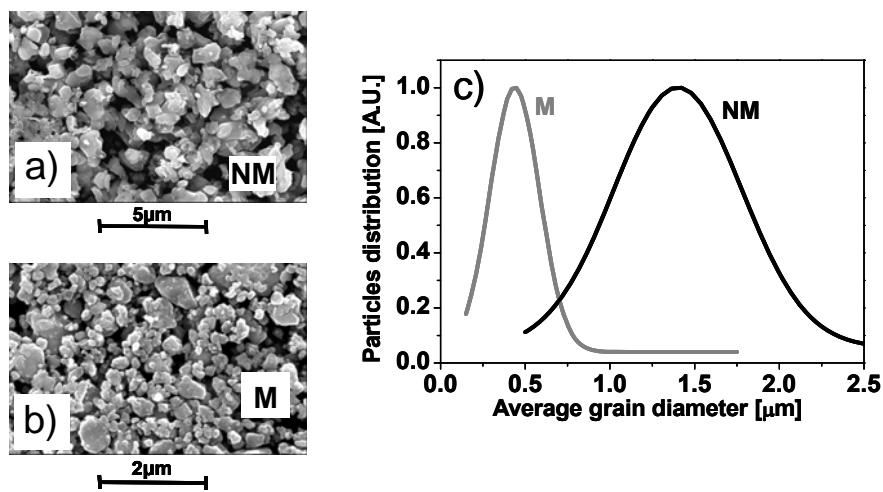

**Figure 1.** SEM images of the not milled (1a) and milled (1b) powders. In 1c) the Gaussian fit of particle size distribution is reported after measuring about 500 grains for each sample. The average diameter is 1.5 μm and 450 nm respectively for the NM and M powders.

In table I a summary of the prepared samples and their geometrical characteristics is reported while in figure 2 the cross sections of the wires are shown. All these multifilamentary samples were heat treated at 980 °C for 3 minutes using a continuous heat treatment system.

Transport critical current ($I_c$) measurements were performed over ~10 cm long samples at the Grenoble High Magnetic Field Laboratory (GHMFL) at 4.2 K in magnetic field up to 13 T, the current was applied perpendicular to the magnetic field. The criterion for the $I_c$ definition was 1 µV/cm.

**Table I**. Summary of the prepared samples and their geometrical characteristics

| sample | N° filaments | powder | Filling factor % | Average filament size [µm] | External shape |
|---|---|---|---|---|---|
| **19NM** | 19 | Not milled | 33 | 270 | square |
| **19M** | 19 | milled | 33 | 270 | square |
| **91NM** | 91 | Not milled | 30 | 110 | square |
| **91M** | 91 | milled | 30 | 110 | square |
| **361NM** | 361 | Not milled | 15 | 30 | square |
| **361M** | 361 | milled | 15 | 30 | square |
| **91Mr** | 91 | milled | 30 | 105 | round |
| **361Mr** | 361 | milled | 14 | 30 | round |

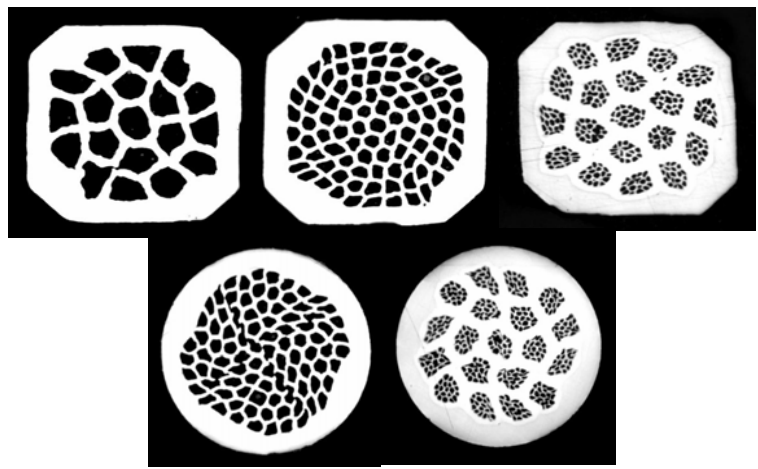

**Figure 2.** Cross sections of the unsintered wires described in table I

## 3. Results and discussion

In figure 3 details of the filaments inside the 91 and 361 multifilamentary square wires after the heat treatment are shown. By SEM analysis the interface between $MgB_2$ and Ni sheath has been investigated giving special attention to the comparison between the milled and not milled samples. As reported in [19], a $MgB_2Ni_{2.5}$ reaction layer appears during the final heat treatment depending on temperature and time. As it was expected, the milled powders are slightly more reactive, nevertheless the undesired layer does not exceed 8 μm: the $MgB_2$ homogeneity is preserved even in the samples with the thinner filaments.

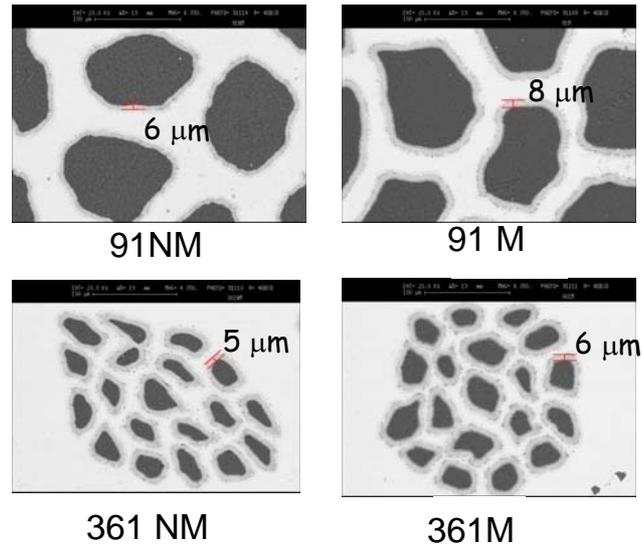

F**igure**. **3**. Details of the filaments inside the 91 and 361 multifilamentary wires with NM and M powders after the heat treatment performed at 980 °C for 3 minutes: the undesired layer does not exceed 8 μm.

The measured critical current densities for all the square samples are reported in figure 4, while in figure 5 the $J_C$ values at 5.5 T are shown.

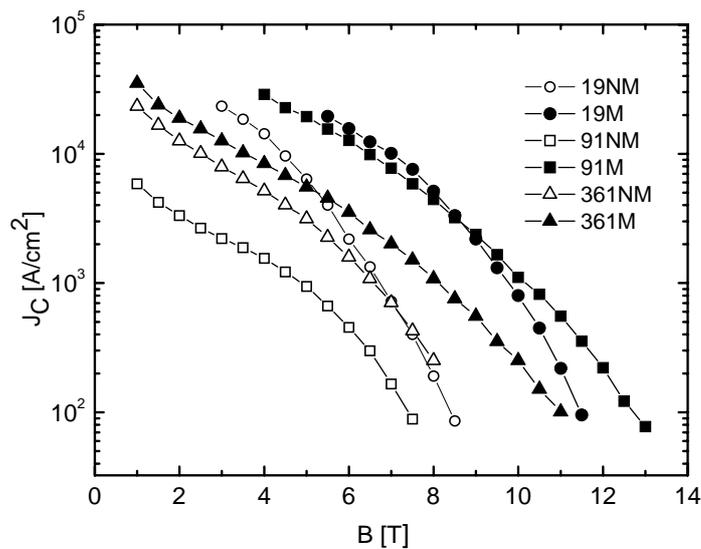

**Figure 4**. Transport critical current density as measured up to 13 T at GHMFL in He bath for all samples.

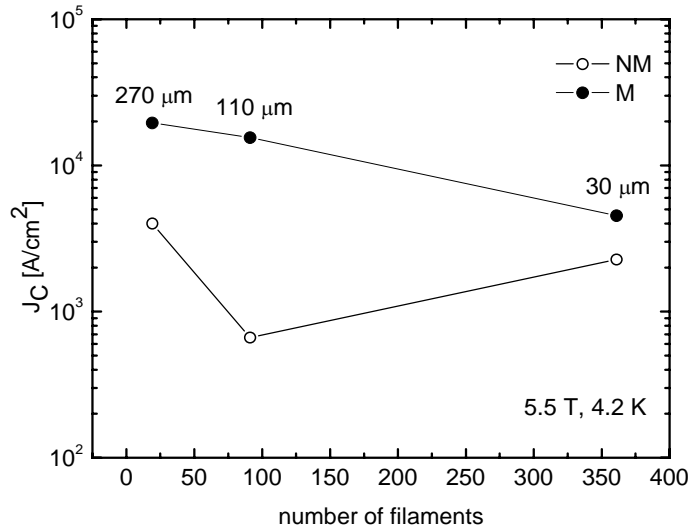

**Figure 5.** Transport critical current density values at 5.5 T and 4.2 K as a function of the filament number for both NM and M samples. The average single filament size for each wire is also reported.

$J_C$ improves with milling for all samples as already described in our previous work [18]. Focusing on the behaviour of the not milled samples, passing from 19 to 91 filaments a remarkable $J_C$ degradation is evident, that is partially recovered going to 361 filaments. On the contrary for the milled samples 19M and 91M have almost identical $J_C$ – a slight better behaviour in field being observed in 91M. When the number of filaments increases up to 361, $J_C$ decreases staying though above the 361NM.

However such $J_C$ behaviours in field can not be explained and well understood simply with the milling effects. Actually in these complex conductors several factors have to be taken into account which have an effect on the transport properties: the starting granulometry of the $MgB_2$ powders, the cold deformation force and the final filament size. As already observed [18] in the *ex-situ* PIT technique the strong cold deformation - besides compacting the powders inside the sheath and thus improving the connectivity - has a crashing effect and further reduces their average grain size. But on the other side, when it is particularly hard the cold deformation could cause micro-cracks or filaments sausaging which act as obstacles on the percolative current paths. In this context it is clear that the flow of the $MgB_2$ grains or agglomerates inside the sheath during the drawing or rolling depends also on the starting granulometry and final single filament size. Therefore, in the final analysis, the capability of these conductors to transport high critical currents crucially depends on a proper balance of these parameters.

Considering what just described, the $J_C$ of the 91NM is considerably lower than the 19NM one because being the starting powder grains size large with respect to the final filament size, a bad grains flow inside the sheath has brought about micro-cracks. On the other side the stronger cold deformation undergone by the 361NM wire – thanks to the double restacking – brought about a higher powder compaction and therefore a partial recover of the $J_C$. This still remains limited by the micro-cracks and/or filament sausaging inevitably occurring in obtaining such thin filaments.

Analysing the behaviour of the milled samples it seems that the $J_C$ enhancement in the sample with 91 filaments is due just to the finer grains of the used powder. For the 361M wire we still observe a $J_C$ improvement with respect to the 361NM thanks to the finer grains, even if the effect is again limited by the micro-cracks: thinner the filaments, more obstructive are the cracks on the percolative current paths.

The reported $J_C$ were calculated for samples having different superconducting sections: to have an idea of the actual current capabilities of the conductors, in figure 6 the corresponding $I_C$ for the wires prepared with the milled powders are shown.

Moreover, to meet the industrial needs for AC applications employing such $MgB_2$ conductors, the best performing conductors underwent a further special cold deformation step to be made round shaped. In figure 7 the $J_C$ of 91Mr and 361Mr (round) samples with 91 and 361 filaments are reported and compared with the corresponding square wires. No significant degradation of the transport properties is observed even after this further deformation step, making them really appealing for AC use.

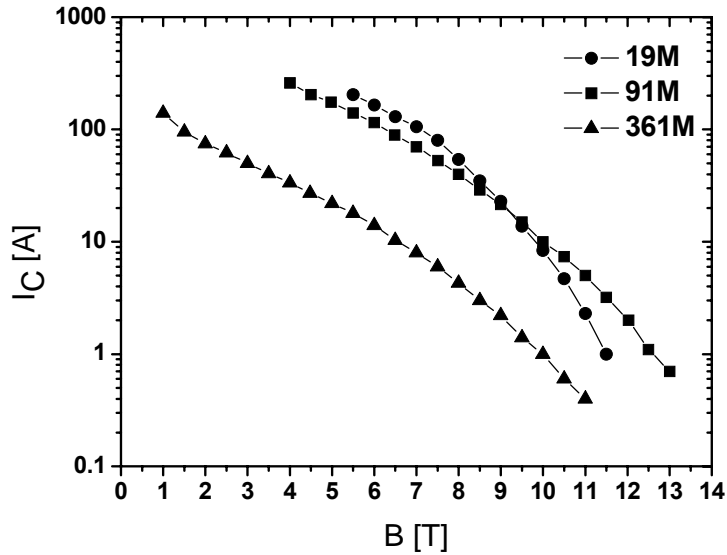

**Figure 6.**. The critical current $I_C$ as a function of the magnetic field for the square wires prepared with the milled powders.

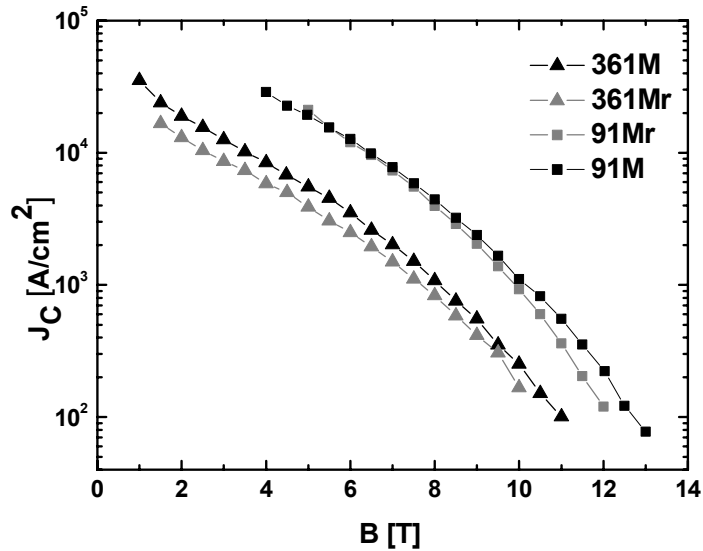

**Figure 7.** Comparison between the $J_C$ of the 91Mr and 361Mr (round shaped) samples and the corresponding square wires. No significant degradation of the transport properties is observed even after the further special cold deformation step.

The measured *I-V* characteristics, through the analysis of the so-called n-value - i.e. the exponential value of the voltage-current characteristic - gave us the possibility to extract information also on the homogeneity of the samples in order to find a better correlation between granulometry, number of filaments and their dimension. In fact, a high n-value above 30 is generally interpreted as a sign of a highly homogeneous superconductor with rather strong pinning force. On the contrary, a low n-value is due to different reasons as a wide spread of the critical current distribution inside the superconducting filaments, typically caused by filament sausaging, impurities, voids and micro-cracks. Furthermore, from an applicative point of view, a high n-value is indicative of the capability of sustaining relevant persistent currents for long periods of time and thus of the possibility of employing a $MgB_2$ conductor to realize windings that can be operated in persistent mode.

The exponential n-value was calculated by plotting the voltage-current characteristics in a double-logarithmic scale and by best fitting with a linear behaviour of the decade in voltage data centred around 1 µV/cm (0.3 to 3 µV/cm). A good linearity of the data plotted in double-logarithmic scale over this voltage range was generally observed, as can be seen in figure 8, where the experimental data and the exponential fit are reported for the 91M sample at a magnetic field of 7 T. The n-values for both NM and M square wire series are reported in figure 9, and they are quite field dependent. Depending on the sample, below a certain field the n-value was hardly quantifiable due to the lack of experimental points caused by very sharp transitions. Both in 19 and 91 filaments samples the n-value increases with milling staying above 30 for magnetic fields up to 8 T and 7.5 T respectively. This is consistent with the fact that after milling the number of grain boundaries, which are pinning centers, is increased and the homogeneity of the powder is improved[18].

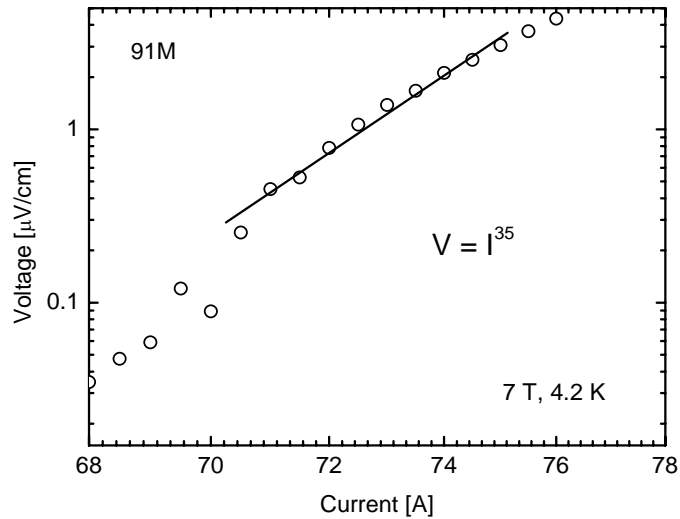

**Figure 8.**. Voltage-Current characteristic and exponential fit plotted in double-logarithmic scale for the 91M sample at 7 T. A linear behaviour of the voltage data is observed in a decade centered around 1 µV/cm.

Concerning the 361 filaments wire the n-values are lower and about the same in NM and M samples. The corresponding decreased homogeneity is consistent with the occurring of micro-cracks or filament sausaging which keep up also in the milled sample. These last results together with the $J_C$ behaviour of the 361NM/M samples suggest that to achieve higher $J_C$ values in so thin filaments it is needed to further decrease the granulometry of the starting powders.

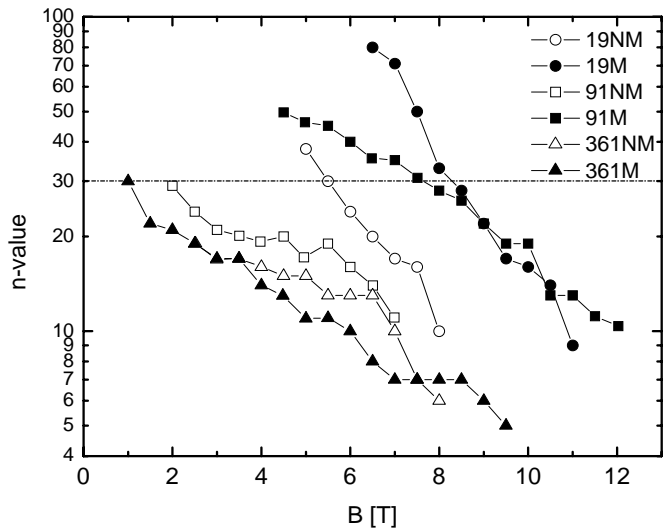

**Figure 9.** The exponential n-values as a function of the magnetic field for both NM and M square wire series.

## 4. Conclusions

Multifilamentary wires with a number of filament up to 361 and an average filament size down to 30 μm have been realized by the *ex-situ* PIT method using pure $MgB_2$ powders with different granulometry. By decreasing the filament size a dependence of the transport properties and the homogeneity on the powder grain size has been observed.

In this work we have obtained the best ratio filament size/grain size on a 91 filaments wire with an average filament size of about 110 μm and a powder starting average grain diameter of about 450 nm. The obtained $J_C$ was $10^4$ A/cm$^2$ at 7 T and 4.2 K. The same results have been obtained on round shape wires which are more suitable for industrial applications.

A finer $MgB_2$ granulometry seems to be needed to realize very thin filaments (10-30 μm) with high critical current density, although the critical current measured on the 361 sample together with the advantages given by the high number of 30 μm thin filaments could be interesting for low field AC applications.

**Acknowledgements**
The authors wish to thank Dr. Eric Mossang for assistance at GHMFL.
The support of Columbus Superconductors S.p.A., "Compagnia di S. Paolo" and European Commission from the 7th Framework Programme Capacities "Transnational Access", Contract N° 228043-EuroMagNET II - Integrated Activities are acknowledged.